\documentclass[11pt]{article}

\usepackage{mathptmx}

\font\twlgot =eufm10 scaled \magstep1 \font\egtgot =eufm8
\font\sevgot =eufm7 \font\twlmsb =msbm10 scaled \magstep1
\font\egtmsb =msbm8 \font\sevmsb =msbm7

\newfam\gotfam
\def\pgot{\fam\gotfam\twlgot}
\textfont\gotfam\twlgot \scriptfont\gotfam\egtgot
\scriptscriptfont\gotfam\sevgot
\def\got{\protect\pgot}
\newfam\msbfam
\textfont\msbfam\twlmsb \scriptfont\msbfam\egtmsb
\scriptscriptfont\msbfam\sevmsb
\def\Bbb{\protect\pBbb}
\def\pBbb{\relax\ifmmode\expandafter\Bb\else\typeout{You cann't use
Bbb in text mode}\fi}
\def\Bb #1{{\fam\msbfam\relax#1}}

\newcommand{\gL}{{\got L}}

\def\thebibliography#1{\section*{References}\list
  {[\arabic{enumi}]}{\settowidth\labelwidth{#1}\leftmargin\labelwidth
    \advance\leftmargin\labelsep
    \usecounter{enumi}}
    \def\newblock{\hskip .11em plus .33em minus .07em}
    \sloppy\clubpenalty4000\widowpenalty4000
    \sfcode`\.=1000\relax}

\let\Large=\large

\def\op#1{\mathop{\fam0 #1}\limits}

\newcommand{\im}{{\rm Im\,}}

\newcommand{\beq}{\begin{equation}}
\newcommand{\eeq}{\end{equation}}
\newcommand{\ben}{\begin{eqnarray}}
\newcommand{\een}{\end{eqnarray}}
\newcommand{\be}{\begin{eqnarray*}}
\newcommand{\ee}{\end{eqnarray*}}
\newcommand{\bea}{\begin{eqalph}}
\newcommand{\eea}{\end{eqalph}}

\newcommand{\cA}{{\cal A}}

\newcommand{\cP}{{\cal P}}

\newcommand{\cL}{{\cal L}}

\newcommand{\cE}{{\cal E}}

\newcommand{\cF}{{\cal F}}
\newcommand{\cS}{{\cal S}}
\newcommand{\cC}{{\cal C}}
\newcommand{\cO}{{\cal O}}

\newcommand{\cG}{{\cal G}}
\newcommand{\bL}{{\bf L}}

\newcommand{\al}{\alpha}

\newcommand{\bt}{\beta}
\newcommand{\dl}{\delta}
\newcommand{\la}{\lambda}
\newcommand{\La}{\Lambda}
\newcommand{\f}{\phi}

\newcommand{\m}{\mu}

\newcommand{\thh}{\theta}

\newcommand{\vt}{\vartheta}

\newcommand{\up}{\upsilon}

\newcommand{\si}{\sigma}
\newcommand{\Si}{\Sigma}
\newcommand{\w}{\wedge}
\newcommand{\wt}{\widetilde}

\newcommand{\ol}{\overline}
\newcommand{\dr}{\partial}
\newcommand{\ar}{\op\longrightarrow}
\newcommand{\ot}{\otimes}

\newcommand{\ve}{\varepsilon}

\newcommand{\rdr}{\stackrel{\leftarrow}{\dr}{}}
\newcommand{\lto}{\leftarrow}
\newcommand{\llr}{\op\longleftarrow}

\newcounter{eqalph}
\newcounter{equationa}
\newcounter{remark}
\newcounter{theorem}
\newcounter{proposition}
\newcounter{lemma}
\newcounter{corollary}
\newcounter{definition}

\def\theremark{\arabic{remark}}

\def\thedefinition{\arabic{definition}}

\newenvironment{proof}{\noindent {\it Proof.}}{}

\newenvironment{theo}{\refstepcounter{definition} \medskip\noindent
THEOREM \thedefinition.\it}{\medskip }

\newenvironment{cor}{\refstepcounter{definition} \medskip\noindent
COROLLARY \thedefinition.\it }{\medskip}
\newenvironment{defi}{\refstepcounter{definition}\medskip \noindent
DEFINITION \thedefinition.\it }{\medskip }

\newenvironment{eqalph}{\stepcounter{equation}
\setcounter{equationa}{\value{equation}} \setcounter{equation}{0}

\begin{eqnarray}}{\end{eqnarray}\setcounter{equation}{\value{equationa}}}

\newcommand{\mar}[1]{}

\begin{document}

{\parindent=0pt

{ \Large \bf The KT-BRST Complex of a Degenerate Lagrangian
System}
\bigskip

{\sc D. BASHKIROV$^1$, G. GIACHETTA$^2$, L. MANGIAROTTI$^2$ and \\
G.SARDANASHVILY$^1$}

\bigskip

{ \small

{\it $^1$ Department of Theoretical Physics, Moscow State
University, 117234 Moscow, Russia
\medskip

$^2$ Department of Mathematics and Informatics, University of
Camerino, 62032 Camerino (MC), Italy}

\bigskip

{\bf Abstract.} Quantization of a Lagrangian field system
essentially depends on its degeneracy and implies its BRST
extension defined by sets of non-trivial Noether  and higher-stage
Noether identities. However, one meets a problem how to select
trivial and non-trivial higher-stage Noether identities. We show
that, under certain conditions, one can associate to a degenerate
Lagrangian $L$ the KT-BRST complex of fields, antifields and
ghosts whose boundary and coboundary operators provide all
non-trivial Noether identities and gauge symmetries of $L$. In
this case, $L$ can be extended to a proper solution of the master
equation.


70S05, 70S20



} }

\section{Introduction}

The BV quantization of Lagrangian field system essentially depends
on its degeneracy and implies its BRST extension given by the
Koszul--Tate (henceforth KT) and BRST complexes
\cite{barn,bat,fisch,gom}. These complexes are defined by sets of
non-trivial Noether identities (henceforth NI) and higher-stage
NI. Any Euler--Lagrange operator satisfies NI which are separated
into the trivial and non-trivial ones. These NI obey first-stage
NI, which in turn are subject to the second-stage NI, and so on.
One however meets a problem how to select trivial and non-trivial
higher-stage NI.

Note that the notion of higher-stage NI has come from that of
reducible constraints. The KT complex of NI has been invented
similarly to that of constraints under the condition that NI are
locally separated into independent and dependent ones
\cite{barn,fisch}. This condition is relevant for constraints,
defined by a finite set of functions which the inverse mapping
theorem is applied to. However, NI unlike constraints are
differential equations. They are given by an infinite set of
functions on a Fr\'echet manifold of infinite order jets where the
inverse mapping theorem fails to be valid.

We consider a generic Lagrangian theory of even and odd variables
on an $n$-dimensional smooth real manifold $X$. It is described in
terms of the Grassmann-graded variational bicomplex
\cite{barn,jmp05,jmp05a,cmp04}. Accordingly, NI are represented by
one-cycles of a certain chain complex. Its boundaries are
necessarily NI, called trivial. Non-trivial NI modulo the trivial
ones are given by first homology of this complex. To describe
$(k+1)$-stage NI, let us assume that non-trivial $k$-stage NI are
generated by a projective $C^\infty(X)$-module $\cC_{(k)}$ of
finite rank and that a certain homology condition (Definition
\ref{v155}) holds \cite{jmp05a}. In this case, $(k+1)$-stage NI
are represented by $(k+2)$-cycles of some chain complex of modules
of antifields isomorphic to $\cC_{(i)}$, $i\leq k$. Accordingly,
trivial $(k+1)$-stage NI are defined as its boundaries. Iterating
the arguments, we come to the exact KT complex (\ref{v94}) with
the boundary KT operator (\ref{v92}) whose nilpotentness is
equivalent to all non-trivial NI (Theorem \ref{t4}) \cite{jmp05a}.

The inverse second Noether theorem (Theorem \ref{w35}) that we
prove associates to the KT complex (\ref{v94}) the cochain
sequence (\ref{w108}) the $(\op\w^nT^*X)$-duals of the modules
$\cC_{(k)}$ whose elements are called ghosts. Components of its
ascent operator (\ref{w108'}), called the gauge operator, are
gauge and higher-stage gauge symmetries of an original Lagrangian
$L$. The gauge operator need not be nilpotent. We show that, if it
admits a nilpotent extension (\ref{w109}), an original Lagrangian
$L$ is extended to a proper solution of the master equation
(\ref{w44}), which the BV quantization starts with. In this case,
the KT complex (\ref{v94}) and the cochain sequence (\ref{w108})
are combined into the KT-BRST complex which is a desired BRST
extension of an original Lagrangian field theory.

\section{Lagrangian theory of even and odd fields}

Let us consider a composite bundle $F\to Y\to X$ where $F\to Y$ is
a vector bundle provided with bundle coordinates $(x^\la, y^i,
q^a)$. Jet manifolds $J^rF$ of $F\to X$ are also vector bundles
$J^rF\to J^rY$ coordinated by $(x^\la, y^i_\La, q^a_\La)$, $0\leq
|\La|\leq r$, with respect to linear frames $\{e^\La_a\}$, where
$\La=(\la_1...\la_k)$, $|\La|=k$, denote symmetric multi-indices.
Let $(J^rY,\cA_r)$ be a graded manifold whose body is $J^rY$ and
whose $C^\infty(J^rY)$-ring of graded functions $\cA_r$ is
generated by sections of the dual $(J^rF)^*$ of $J^rF\to J^rY$,
i.e, it is locally generated by the coframes $\{c^a_\La\}$ dual of
the frames $\{e_a^\La\}$. Let $\cS^*_r[F;Y]$ be the differential
graded algebra (henceforth DGA) of graded differential forms on
the graded manifold $(J^rY,\cA_r)$. There is the inverse system of
jet manifolds $J^{r-1}Y \leftarrow J^rY$. Its projective limit is
a Fr\'echet manifold coordinated by $(x^\la, y^i_\La)$,
$0\leq|\La|$. This inverse system yields the direct system of DGAs
\be
\cS^*[F;Y]\ar \cS^*_1[F;Y]\ar\cdots \cS^*_r[F;Y]\ar\cdots.
\ee
Its direct limit $\cS^*_\infty[F;Y]$ is the DGA of all graded
differential forms on graded manifolds $(J^rY,\cA_r)$.  It
contains the subalgebra $\cO^*_\infty Y$ of all exterior forms on
jet manifolds $J^rY$. It is an $\cO^0_\infty Y$-algebra locally
generated by elements $(c^a_\La,dx^\la,dy^i_\La, dc^a_\La)$,
$0\leq |\La|$. The collective symbol $(s^A)$ further stands for
the tuple $(y^i,c^a)$, called a local basis for the DGA
$\cS^*_\infty[F;Y]$. Let $[A]=[s^A]$ denote the Grassmann parity.

With this notation, a graded derivation of the $\Bbb R$-ring
$\cS^0_\infty[F;Y]$ takes the form
\mar{gg}\beq
 \vt=\vt^\la\dr_\la + \op\sum_{0\leq|\La|} \vt_\La^A\dr^\La_A,
 \qquad \dr^\La_A(s_\Si^B)=\dr^\La_A\rfloor
ds_\Si^B=\dl_A^B\dl^\La_\Si. \label{gg}
\eeq
It yields the Lie derivative $\bL_\vt\f=\vt\rfloor d\f+
d(\vt\rfloor\f)$ of the DGA $\cS^*_\infty[F;Y]$. In particular,
the total derivatives are defined as the derivations
\be
d_\la =\dr_\la + \op\sum_{0\leq|\La|} s_{\la+\La}^A\dr^\La_A,
\qquad d_\La=d_{\la_1}\cdots d_{\la_k}, \qquad \la
+\La=(\la\la_1...\la_k).
\ee

The DGA $\cS^*_\infty[F;Y]$ is split into the Grassmann-graded
variational bicomplex of modules $\cS^{k,r}_\infty[F;Y]$ of
$r$-horizontal and $k$-contact graded forms locally generated by
one-forms $dx^\la$ and $\thh^A_\La=ds_\La^A -s^A_{\la +\La}
dx^\la$ \cite{barn,jmp05,jmp05a,cmp04}. It contains the
variational subcomplex
\mar{g111}\beq
0\to \Bbb R\ar
\cS^0_\infty[F;Y]\ar^{d_H}\cS^{0,1}_\infty[F;Y] \cdots \ar^{d_H}
\cS^{0,n}_\infty[F;Y]\ar^\dl \cS^{1,n}_\infty[F;Y], \label{g111}
\eeq
where $d_H(\f)=dx^\la\w\bL_{d_\la}(\f)$ is the total
differential and $\dl$ is the variational operator. Lagrangians
and Euler--Lagrange operators are defined as its even elements
\mar{0709}\ben
&& L=\cL d^nx\in \cS^{0,n}_\infty[F;Y], \nonumber\\
&&\dl L= \thh^A\w \cE_A d^nx=\op\sum_{0\leq|\La|}
 (-1)^{|\La|}\thh^A\w d_\La (\dr^\La_A \cL) d^nx, \label{0709}
\een

The relevant cohomology of the variational bicomplex has been
obtained \cite{cmp04,epr05}. In particular, any variationally
trivial ($\dl$-closed) odd density $L\in \cS^{0,n}_\infty[F;Y]$ is
$d_H$-exact, and the form $dL-\dl L$ is $d_H$-exact for any
density $L\in \cS^{0,n}_\infty[F;Y]$.

A graded derivation $\vt$ (\ref{gg}) is called a variational
symmetry of a Lagrangian $L$ if the Lie derivative $\bL_\vt L$ is
$d_H$-exact. We restrict our consideration to vertical contact
graded derivations $\vt$ vanishing on $C^\infty(X)\subset
\cS^0_\infty[F;Y]$ and preserving the ideal of contact forms of
the DGA $\cS^*_\infty[F;Y]$. Such a derivation is the jet
prolongation
\mar{0672}\beq
\vt=\up^A\dr_A + \op\sum_{0<|\La|}
d_\La\up^A\dr_A^\La \label{0672}
\eeq
of its restriction $\up=\up^A\dr_A$ to the ring
$\cS^0_\infty[F;Y]$. It obeys the relations $\vt\rfloor
d_H\f=-d_H(\vt\rfloor\f)$, $\f\in\cS^*_\infty[F;Y]$. In
particular, we have
\mar{xx10}\beq
\bL_{\vt}L= \up\rfloor\dl L
+d_H\si=\up^A\cE_A d^nx+d_H\si \label{xx10}
\eeq
for any Lagrangian $L$. It follows that $\vt$ (\ref{0672}) is a
variational symmetry of $L$ iff the form $\up\rfloor\dl L$ is
$d_H$-exact. A graded derivation $\vt$ (\ref{0672}) is called
nilpotent if $\bL_\vt(\bL_\vt\f)=0$ for any horizontal form $\f\in
\cS^{0,*}_\infty[F;Y]$. It is nilpotent only if it is odd and iff
$\vt(\up)=0$.

For the sake of simplicity, the common symbol $\up$ further stands
for the graded derivation $\vt$ (\ref{0672}), its summand $\up$,
and the Lie derivative $\bL_\vt$. We agree to call $\up$ the
graded derivation of the DGA $\cS^*_\infty[F;Y]$. Its right graded
derivations $\op\up^\lto ={\op\dr^\lto}_A\up^A$ are also
considered. One associates to any right graded derivation
$\op\up^\lto ={\op\dr^\lto}_A\up^A$ the left one
\be
\up^l=(-1)^{[\up][A]}\up^A\dr_A, \qquad
\up^l(f)=(-1)^{[\up][f]}\up(f), \qquad f\in \cS^0_\infty[F;Y].
\ee

\section{KT complex of Noether identities}

Let us start with the following notation. Given a vector bundle
$E\to X$, we call $\ol E=E^*\ot\op\w^n T^*X$ the density-dual of
$E$. The density dual of a graded vector bundle $E=E^0\oplus E^1$
is $\ol E=\ol E^1\oplus \ol E^0$. Given a graded vector bundle
$E=E^0\oplus E^1$ over $Y$, we consider the composite bundle $E\to
E^0\to X$ and denote $\cP^*_\infty[E;Y]=\cS^*_\infty[E;E^0]$.

Given a Lagrangian theory $(\cS^*_\infty[F;Y],L)$, its
Euler--Lagrange operator $\dl L$ (\ref{0709}) obeys NI defined by
linear differential operators on the $\cS^0_\infty[F;Y]$-module
$\cS^{1,n}_\infty[F;Y]$ whose kernels contain $\dl L$. To describe
these NI let us consider the density-dual
$\ol{VF}=V^*F\ot_F\op\w^n T^*X$ of the vertical tangent bundle
$VF\to F$. We assume that $F\to Y$ is a trivial vector bundle. In
this case,
\be
\ol{VF}=(\ol F\oplus V^*Y\op\ot_Y\op\w^n T^*X)\oplus F
\ee
is a graded vector bundle over $Y$. Let us enlarge
$\cS^*_\infty[F;Y]$ to the DGA $\cP^*_\infty[\ol{VF};Y]$ with a
local basis $(s^A, \ol s_A)$, $[\ol s_A]=([A]+1){\rm mod}\,2$. Its
elements $\ol s_A$ are called antifields of antifield number
Ant$[\ol s_A]= 1$. The DGA $\cP^*_\infty[\ol{VF};Y]$ is endowed
with the nilpotent right graded derivation $\ol\dl=\rdr^A \cE_A$.
Then we have the chain complex
\mar{v042}\beq
0\lto \im\ol\dl \llr^{\ol\dl} \cP^{0,n}_\infty[\ol{VF};Y]_1
\llr^{\ol\dl} \cP^{0,n}_\infty[\ol{VF};Y]_2 \label{v042}
\eeq
of graded densities of antifield number $\leq 2$. Its one-chains
are linear differential operators on $\cS^{1,n}_\infty[F;Y]$, and
its one-cycles define the NI
\mar{0712,3}\ben
&& \ol\dl \Phi=0, \qquad \Phi= \op\sum_{0\leq|\La|}
\Phi^{A,\La}\ol s_{\La A} d^nx \in \cP^{0,n}_\infty[\ol{VF};Y]_1,
\label{0712}\\
&& \op\sum_{0\leq|\La|} \Phi^{A,\La} d_\La \cE_A d^nx=0.
\label{0713}
\een
Conversely, all NI (\ref{0713}) come from the cycles (\ref{0712}).
In particular, one-chains $\Phi \in \cP^{0,n}_\infty[\ol{VF};Y]_1$
are necessarily NI if they are boundaries. Therefore, these NI are
called trivial. Accordingly, non-trivial NI modulo the trivial
ones correspond to elements of the first homology $H_1(\ol\dl)$ of
the complex (\ref{v042}) \cite{jmp05a}. A Lagrangian $L$ is called
degenerate if there are non-trivial NI.

Non-trivial NI obey first-stage NI. To describe them, let us
assume that the module $H_1(\ol \dl)$ is finitely generated.
Namely, there exists a projective $C^\infty(X)$-module
$\cC_{(0)}\subset H_1(\ol \dl)$ of finite rank possessing a local
basis $\{\Delta_r\}$ such that any element $\Phi\in H_1(\ol \dl)$
factorizes
\mar{xx2}\beq
\Phi= \op\sum_{0\leq|\Xi|} G^{r,\Xi} d_\Xi \Delta_r d^nx, \qquad
\Delta_r=\op\sum_{0\leq|\La|} \Delta_r^{A,\La}\ol s_{\La A},\qquad
G^{r,\Xi},\Delta_r^{A,\La}\in \cS^0_\infty[F;Y], \label{xx2}
\eeq
via elements of $\cC_{(0)}$. Thus, all non-trivial NI (\ref{0713})
result from the NI
\mar{v64}\beq
\ol\dl\Delta_r= \op\sum_{0\leq|\La|} \Delta_r^{A,\La} d_\La
\cE_A=0. \label{v64}
\eeq
By virtue of the Serre--Swan theorem, the module $\cC_{(0)}$ is
isomorphic to a module of sections of the density-dual $\ol E_0$
of some graded vector bundle $E_0\to X$. Let us enlarge
$\cP^*_\infty[\ol{VF};Y]$ to the DGA
$\ol\cP^*_\infty\{0\}=\cP^*_\infty[\ol{VF}\oplus_Y \ol E_0;Y]$
possessing a local basis $(s^A,\ol s_A, \ol c_r)$ of Grassmann
parity $[\ol c_r]=([\Delta_r]+1){\rm mod}\,2$ and antifield number
${\rm Ant}[\ol c_r]=2$. This DGA is provided with the odd right
graded derivation $\dl_0=\ol\dl + \rdr^r\Delta_r$ which is
nilpotent iff the NI (\ref{v64}) hold. Then we have the chain
complex
\mar{v66}\beq
0\lto \im\ol\dl \op\lto^{\ol\dl}
\cP^{0,n}_\infty[\ol{VF};Y]_1\op\lto^{\dl_0}
\ol\cP^{0,n}_\infty\{0\}_2 \op\lto^{\dl_0}
\ol\cP^{0,n}_\infty\{0\}_3 \label{v66}
\eeq
of graded densities of antifield number $\leq 3$. It possesses
trivial homology $H_0(\dl_0)$ and $H_1(\dl_0)$. Its two-cycles
define the first-stage NI
\mar{v79}\ben
&& \dl_0 \Phi=0, \qquad \Phi= G + H= \op\sum_{0\leq|\La|} G^{r,\La}\ol c_{\La r}d^nx +
\op\sum_{0\leq|\La|,|\Si|} H^{(A,\La)(B,\Si)}\ol s_{\La A}\ol
s_{\Si B}d^nx,  \nonumber\\
&& \op\sum_{0\leq|\La|} G^{r,\La}d_\La\Delta_r d^nx =-\ol\dl H.
\label{v79}
\een
However, the converse need not be true. One can show that NI
(\ref{v79}) are cycles iff any $\ol\dl$-cycle $\Phi\in
\cP^{0,n}_\infty[\ol{VF};Y]_2$ is a $\dl_0$-boundary
\cite{jmp05a}. In particular, a cycle $\Phi$ is a boundary if its
summand $G$ is $\ol\dl$-exact. Any boundary $\Phi\in
\ol\cP^{0,n}_\infty\{0\}_2$ necessarily defines first-stage NI
(\ref{v79}), called trivial. Accordingly, non-trivial first-stage
NI modulo the trivial ones are identified to elements of the
second homology $H_2(\dl_0)$ of the complex (\ref{v66}). Note that
this definition is independent on specification of a generating
module $\cC_{(0)}$. Given a different one, there exists a chain
isomorphism between the corresponding complexes (\ref{v66}).

A degenerate Lagrangian is called reducible if there are
non-trivial first-stage NI. These obey second-stage NI, and so on.
Iterating the arguments, we say that a degenerate Lagrangian is
$N$-stage reducible if the following hold \cite{jmp05a}.

(i) There are graded vector bundles $E_0,\ldots, E_N$ over $X$,
and the DGA $\cP^*_\infty[\ol{VF};Y]$ is enlarged to the DGA
\mar{v91}\beq
\ol\cP^*_\infty\{N\}=\cP^*_\infty[\ol{VF}\op\oplus_Y \ol
E_0\op\oplus_Y\cdots\op\oplus_Y \ol E_N;Y] \label{v91}
\eeq
with a local basis $(s^A,\ol s_A, \ol c_r, \ol c_{r_1}, \ldots,
\ol c_{r_N})$ of antifield number Ant$[\ol c_{r_k}]=k+2$.

(ii) The DGA (\ref{v91}) is provided with the nilpotent right
graded derivation
\mar{v92,'}\ben
&&\dl_{KT}=\rdr^A\cE_A +
\op\sum_{0\leq|\La|}\rdr^r\Delta_r^{A,\La}\ol s_{\La A} +
\op\sum_{1\leq k\leq N}\rdr^{r_k} \Delta_{r_k},
\label{v92}\\
&& \Delta_{r_k}= \op\sum_{0\leq|\La|}
\Delta_{r_k}^{r_{k-1},\La}\ol c_{\La r_{k-1}} + \op\sum_{0\leq
|\Si|, |\Xi|}(h_{r_k}^{(r_{k-2},\Si)(A,\Xi)}\ol c_{\Si r_{k-2}}\ol
s_{\Xi A}+...), \label{v92'}
\een
of antifield number -1, where the index $k=-1$ stands for $\ol
s_A$.

(iii) The module $\ol\cP^{0,n}_\infty\{N\}_{\leq N+3}$ of
densities of antifield number $\leq (N+3)$ is split into the exact
KT chain complex
\mar{v94}\ben
&& 0\lto \im \ol\dl \llr^{\ol\dl}
\cP^{0,n}_\infty[\ol{VF};Y]_1\llr^{\dl_0}
\ol\cP^{0,n}_\infty\{0\}_2\llr^{\dl_1}
\ol\cP^{0,n}_\infty\{1\}_3\cdots
\label{v94}\\
&& \qquad
 \llr^{\dl_{N-1}} \ol\cP^{0,n}_\infty\{N-1\}_{N+1}
\llr^{\dl_{KT}} \ol\cP^{0,n}_\infty\{N\}_{N+2}\llr^{\dl_{KT}}
\ol\cP^{0,n}_\infty\{N\}_{N+3} \nonumber
\een
which satisfies the following homology condition.

\begin{defi} \label{v155} \mar{v155} One says that
the homology regularity condition holds if any $\dl_{k<N}$-cycle
$\f\in \ol\cP_\infty^{0,n}\{k\}_{k+3}\subset
\ol\cP_\infty^{0,n}\{k+1\}_{k+3}$ is a $\dl_{k+1}$-boundary.
\end{defi}

\begin{theo} \label{t4} \mar{t4} Given an $N$-reducible Lagrangian,
the nilpotentness $\dl_{KT}^2=0$ of the KT operator (\ref{v92}) is
equivalent to the non-trivial NI (\ref{v64}) and $(k\leq N)$-stage
NI
\mar{v93}\beq
\op\sum_{0\leq|\La|} \Delta_{r_k}^{r_{k-1},\La}d_\La
(\op\sum_{0\leq|\Si|} \Delta_{r_{k-1}}^{r_{k-2},\Si}\ol c_{\Si
r_{k-2}}) = -  \ol\dl(\op\sum_{0\leq |\Si|,
|\Xi|}h_{r_k}^{(r_{k-2},\Si)(A,\Xi)}\ol c_{\Si r_{k-2}}\ol s_{\Xi
A}). \label{v93}
\eeq
\end{theo}

\section{Inverse second Noether theorem}

A gauge symmetry of a Lagrangian $L$ is defined as a linear
differential operator on some projective $C^\infty(X)$-module of
finite rank with values in the module of variational symmetries of
$L$. It can be described as follows \cite{jpa05,jmp05}. Let
$E_0\to X$ be a graded vector bundle. Let us enlarge the DGA
$\cS^*_\infty[F;Y]$ to the DGA $\cP^*_\infty[F\oplus_Y E_0;Y]$
possessing a local basis $(s^A,c^r)$ whose elements are called
ghosts. A gauge symmetry of $L$ is an odd vertical contact graded
derivation $u$ of $\cP^0_\infty[F\oplus_Y E_0;Y]$ which is a
variational symmetry of $L$. Any Lagrangian admits gauge
symmetries. Let $u_{(1)}$ be a differential operator on some
projective $C^\infty(X)$-module of finite rank such that $u\circ
u_{(1)}$ is $\ol\dl$-exact. It is called the first-stage gauge
symmetry, and so on.

Different variants of the second Noether theorem relate reducible
NI and gauge symmetries \cite{barn,jpa05,jmp05,fulp}. Given the
DGA $\ol\cP^*_\infty\{N\}$ (\ref{v91}), let us consider the DGA
\mar{w5}\beq
\cP^*_\infty\{N\}=\cP^*_\infty[F\op\oplus_Y E_0\op\oplus_Y\cdots
\op\oplus_Y E_N;Y], \label{w5}
\eeq
possessing a local basis $(s^A, c^r, c^{r_1}, \ldots, c^{r_N})$,
$[c^{r_k}]=([\ol c_{r_k}]+1){\rm mod}\,2$, and the DGA
\mar{w6}\beq
P^*_\infty\{N\}=\cP^*_\infty[\ol{VF}\op\oplus_Y E_0\oplus\cdots
\op\oplus_Y E_N \op\oplus_Y \ol E_0\op\oplus_Y\cdots\op\oplus_Y
\ol E_N;Y]
 \label{w6}
\eeq
with a local basis $(s^A, \ol s^A, c^r, c^{r_1}, \ldots,
c^{r_N},\ol c_r, \ol c_{r_1}, \ldots, \ol c_{r_N})$. Their
elements $c^{r_k}$ are called ghosts of ghost number
gh$[c^{r_k}]=k+1$ and antifield number ${\rm
Ant}[c^{r_k}]=-(k+1)$. The DGAs $\ol\cP^*_\infty\{N\}$ (\ref{v91})
and $\cP^*_\infty\{N\}$ (\ref{w5}) are subalgebras of
$P^*_\infty\{N\}$ (\ref{w6}). The KT operator $\dl_{KT}$
(\ref{v92}) is naturally extended to a graded derivation of the
DGA $P^*_\infty\{N\}$.

\begin{theo} \label{w35} \mar{w35} Given the KT complex (\ref{v94}),
the module of graded densities $\cP_\infty^{0,n}\{N\}$ is split
into the cochain sequence
\mar{w108,'}\ben
&& 0\to \cS^{0,n}_\infty[F;Y]\ar^{u_e}
\cP^{0,n}_\infty\{N\}^1\ar^{u_e}
\cP^{0,n}_\infty\{N\}^2\ar^{u_e}\cdots, \label{w108}\\
&& u_e=u + u_{(1)}+\cdots + u_{(N)}=u^A\frac{\dr}{\dr s^A} +
u^r\frac{\dr}{\dr c^r} +\cdots + u^{r_{N-1}}\frac{\dr}{\dr
c^{r_{N-1}}}, \label{w108'}
\een
graded in the ghost number, where $u$ (\ref{w33}) and $u_{(k)}$
(\ref{w38}), $k=1,\ldots, N$,  are the gauge and higher-stage
gauge symmetries of an original Lagrangian $L$.
\end{theo}

\begin{proof} Note that
any tuple $(f', f^\La)$, $|\La|\leq k$, of local graded functions
$f', f^\La\in \cS^0_\infty[F;Y]$ obeys the relations
\mar{qq1}\ben
&& \op\sum_{0\leq |\La|\leq k} f^\La d_\La f'd^nx= \op\sum_{0\leq
|\La|\leq k}(-1)^{|\La|}d_\La (f^\La)f d^nx +d_H\si,
\label{qq1a}\\
&& \op\sum_{0\leq |\La|\leq k} (-1)^{|\La|}d_\La(f^\La \f)=
\op\sum_{0\leq |\La|\leq k} \eta (f)^\La d_\La \f, \label{qq1b}\\
&& \eta (f)^\La = \op\sum_{0\leq|\Si|\leq k-|\La|}(-1)^{|\Si+\La|}
\frac{(|\Si+\La|)!}{|\Si|!|\La|!} d_\Si f^{\Si+\La}, \quad
(\eta\circ\eta)(f)^\La=f^\La. \label{qq1c}
\een
Let us extend an original Lagrangian $L$ to the Lagrangian
\mar{w8}\beq
L_e=L+L_1=L + \op\sum_{0\leq k\leq N} c^{r_k}\Delta_{r_k}d^nx=L
+\dl_{KT}( \op\sum_{0\leq k\leq N} c^{r_k}\ol c_{r_k}d^nx)
\label{w8}
\eeq
of zero antifield number. It is readily observed that the KT
operator $\dl_{KT}$ is a variational symmetry of $L_e$. It follows
that
\mar{w16}\ben
&& [\frac{\op\dl^\lto \cL_e}{\dl \ol s_A}\cE_A
+\op\sum_{0\leq k\leq N} \frac{\op\dl^\lto \cL_e}{\dl \ol
c_{r_k}}\Delta_{r_k}]d^nx = [\up^A\cE_A + \op\sum_{0\leq k\leq
N}\up^{r_k}\frac{\dl
\cL_e}{\dl c^{r_k}}]d^nx= d_H\si,  \label{w16}\\
&& \up^A= \frac{\op\dl^\lto \cL_e}{\dl \ol s_A}=u^A+w^A
=\op\sum_{0\leq|\La|} c^r_\La\eta(\Delta^A_r)^\La +
 \op\sum_{1\leq i\leq N}\op\sum_{0\leq|\La|}
c^{r_i}_\La\eta(\op\dr^\lto{}^A(h_{r_i}))^\La, \nonumber\\
&& \up^{r_k}=\frac{\op\dl^\lto \cL_e}{\dl \ol c_{r_k}} =u^{r_k}+
w^{r_k}= \op\sum_{0\leq|\La|}
c^{r_{k+1}}_\La\eta(\Delta^{r_k}_{r_{k+1}})^\La +
\op\sum_{k+1<i\leq N} \op\sum_{0\leq|\La|}
c^{r_i}_\La\eta(\op\dr^\lto{}^{r_k}(h_{r_i}))^\La. \nonumber
\een
The equality (\ref{w16}) falls into the set of equalities
\mar{w19,20}\ben
&& \frac{\op\dl^\lto (c^r\Delta_r)}{\dl \ol s_A}\cE_A d^nx
=u^A\cE_A d^nx=d_H\si_0, \label{w19}\\
&&  [\frac{\op\dl^\lto (c^{r_k}\Delta_{r_k})}{\dl \ol s_A}\cE_A
+\op\sum_{0\leq i<k} \frac{\op\dl^\lto (c^{r_k}\Delta_{r_k})}{\dl
\ol c_{r_i}}\Delta_{r_i}] d^nx= d_H\si_k, \qquad k=1,\ldots,N.
\label{w20}
\een
By virtue of the equality (\ref{w19}) and  the formula
(\ref{xx10}), the graded derivation
\mar{w33}\beq
u= u^A\frac{\dr}{\dr s^A}, \qquad u^A =\op\sum_{0\leq|\La|}
c^r_\La\eta(\Delta^A_r)^\La, \label{w33}
\eeq
of $\cP^0\{0\}$ is a variational and, consequently, gauge symmetry
of a Lagrangian $L$ \cite{jmp05,cmp04}. Every equality (\ref{w20})
falls into a set of equalities graded by the polynomial degree in
antifields. Let us consider the equality, linear in antifields
$\ol c_{r_{k-2}}$. We have
\be
&& [\frac{\op\dl^\lto}{\dl \ol
s_A}(c^{r_k}\op\sum_{0\leq|\Si|,|\Xi|}h_{r_k}^{(r_{k-2},\Si)(A,\Xi)}
\ol
c_{\Si r_{k-2}}\ol s_{\Xi A})\cE_A + \\
&& \qquad \frac{\op\dl^\lto}{\dl \ol
c_{r_{k-1}}}(c^{r_k}\op\sum_{0\leq|\Si|}\Delta_{r_k}^{r'_{k-1},\Si}\ol
c_{\Si r'_{k-1}})\op\sum_{0\leq|\Xi|}
\Delta_{r_{k-1}}^{r_{k-2},\Xi}\ol c_{\Xi r_{k-2}}]d^nx= d_H\si_k.
\ee
This equality is brought into the form
\be
 [\op\sum_{0\leq|\Xi|}
(-1)^{|\Xi|}d_\Xi(c^{r_k}\op\sum_{0\leq|\Si|}
h_{r_k}^{(r_{k-2},\Si)(A,\Xi)} \ol c_{\Si r_{k-2}})\cE_A +
u^{r_{k-1}}\op\sum_{0\leq|\Xi|} \Delta_{r_{k-1}}^{r_{k-2},\Xi}\ol
c_{\Xi r_{k-2}}] d^nx= d_H\si_k.
\ee
Using the relation (\ref{qq1a}), we obtain the equality
\be
[\op\sum_{0\leq|\Xi|} c^{r_k}\op\sum_{0\leq|\Si|}
h_{r_k}^{(r_{k-2},\Si)(A,\Xi)} \ol c_{\Si r_{k-2}} d_\Xi\cE_A +
u^{r_{k-1}}\op\sum_{0\leq|\Xi|} \Delta_{r_{k-1}}^{r_{k-2},\Xi}\ol
c_{\Xi r_{ki-2}}]d^nx= d_H\si'_k.
\ee
The variational derivative of both its sides with respect to $\ol
c_{r_{k-2}}$ leads to the relation
\mar{w34}\ben
&&\op\sum_{0\leq|\Si|} d_\Si u^{r_{k-1}}\frac{\dr}{\dr
c^{r_{k-1}}_\Si} u^{r_{k-2}} =\ol\dl(\al^{r_{k-2}}),\label{w34}\\
&& \al^{r_{k-2}} = -\op\sum_{0\leq|\Si|}
\eta(h_{r_k}^{(r_{k-2})(A,\Xi)})^\Si d_\Si(c^{r_k} \ol s_{\Xi A}).
\nonumber
\een
This is the $k$-stage gauge symmetry condition \cite{jmp05}. Thus,
the odd graded derivations
\mar{w38}\beq
u_{(k)}= u^{r_{k-1}}\frac{\dr}{\dr c^{r_{k-1}}}, \qquad
u^{r_{k-1}}=\op\sum_{0\leq|\La|}
c^{r_k}_\La\eta(\Delta^{r_{k-1}}_{r_k})^\La, \qquad k=1,\ldots,N,
\label{w38}
\eeq
are $k$-stage gauge symmetries. Graded derivations $u$
(\ref{w33}), $u_{(k)}$ (\ref{w38}) form the ascent gauge operator
(\ref{w108}) of ghost number 1. It provides the cochain sequence
(\ref{w108}).
\end{proof}

Following the proof of Theorem \ref{w35}, one can show that any
$C^\infty(X)$-module of NI yields a gauge symmetry of a
Lagrangian. Since the gauge operator (\ref{w108'}) need not be
nilpotent, the direct second Noether theorem can not be formulated
in homology terms. Therefore, gauge and higher-stage gauge
symmetries are said to be non-trivial if they are associated to
non-trivial NI and higher-stage NI, respectively.

With the gauge operator (\ref{w108'}), the extended Lagrangian
$L_e$ (\ref{w8}) takes the form
\mar{lmp2}\beq
L_e= L+u_e( \op\sum_{0\leq k\leq N} c^{r_{k-1}}\ol c_{r_{k-1}})
d^nx + L^*_1 + d_H\si, \label{lmp2}
\eeq
where $L^*_1$ is a term of polynomial degree in antifields
exceeding 1.

\section{KT-BRST complex}

The DGA $P^*_\infty\{N\}$ (\ref{w6}) exemplifies a field-antifield
theory of the following type \cite{barn,gom}. Let $Z'\to Z\to X$
be a composite bundle where $Z'\to Z$ is a trivial vector bundle.
Let us consider the DGA $\cP^*_\infty[\ol{VZ'};Z]$ with a local
basis $(z^a,\ol z_a)$, where $[\ol z_a]=([z^a]+1){\rm mod}\,2$.
Its elements $z^a$ and $\ol z_a$ are called fields and antifields,
respectively. Densities of this DGA are endowed with the
antibracket
\mar{f11}\beq
\{\gL d^nx,\gL'd^nx\}=[\frac{\op\dl^\lto \gL}{\dl \ol
z_a}\frac{\dl \gL'}{\dl z^a} +
(-1)^{[\gL']([\gL']+1)}\frac{\op\dl^\lto \gL'}{\dl \ol
z_a}\frac{\dl \gL}{\dl z^a}]d^nx. \label{f11}
\eeq
Furthermore, one associates to any Lagrangian $\gL d^nx$ the odd
graded derivations
\mar{w37,lmp1}\ben
&&\up_\gL=\op\cE^\lto{}^a\dr_a=\frac{\op\dl^\lto \gL}{\dl \ol z_a}
\frac{\dr}{\dr z^a}, \qquad
\ol\up_\gL=\rdr^a\cE_a=\frac{\op\dr^\lto}{\dr \ol z_a}\frac{\dl
\gL}{\dl z^a}, \label{w37}\\
&& \vt_\gL=\up_\gL+ \ol\up_\gL^l=(-1)^{[a]+1}( \frac{\dl \gL}{\dl
\ol z^a}\frac{\dr}{\dr z_a}+\frac{\dl \gL}{\dl z^a}\frac{\dr}{\dr
\ol z_a}),  \label{lmp1}\\
&& \vt_\gL(\gL' d^nx)=\{\gL d^nx,\gL' d^nx\}. \nonumber
\een

\begin{theo} \label{w39} \mar{w39} The following conditions are
equivalent. (i) The antibracket of a Lagrangian $\gL d^nx$ is
$d_H$-exact, i.e.,
\mar{w44}\beq
\{\gL d^nx,\gL d^nx\}=2\frac{\op\dl^\lto \gL}{\dl \ol
z_a}\frac{\dl \gL}{\dl z^a} d^nx =d_H\si. \label{w44}
\eeq
(ii) The graded derivation $\up$ (\ref{w37}) is a variational
symmetry of a Lagrangian $\gL d^nx$. (iii) The graded derivation
$\ol\up$ (\ref{w37}) is a variational symmetry of $\gL d^nx$. (iv)
The graded derivation $\vt_\gL$ (\ref{lmp1}) is nilpotent.
\end{theo}

\begin{proof} By virtue of the formula
(\ref{xx10}), conditions (ii) and (iii) are equivalent to
condition (i). The equality (\ref{w44}) is equivalent to that the
odd density $\op\cE^\lto{}^a\cE_a d^nx$ is variationally trivial.
Replacing right variational derivatives $\op\cE^\lto{}^a$ with
$(-1)^{[a]+1}\cE^a$, we obtain
\be
2\op\sum_a (-1)^{[a]}\cE^a\cE_a d^nx=d_H\si.
\ee
The variational operator acting on this relation results in the
equalities
\be
&&
\op\sum_{0\leq|\La|}(-1)^{[a]+|\La|}d_\La(\dr^\La_b(\cE^a\cE_a))=
\op\sum_{0\leq|\La|}(-1)^{[a]}[\eta(\dr_b\cE^a)^\La\cE_{\La a} +
\eta(\dr_b\cE_a)^\La\cE^a_\La)]=0, \\
&& \op\sum_{0\leq|\La|}(-1)^{[a]+|\La|}d_\La(\dr^{\La
b}(\cE^a\cE_a)) =
\op\sum_{0\leq|\La|}(-1)^{[a]}[\eta(\dr^b\cE^a)^\La\cE_{\La a} +
\eta(\dr^b\cE_a)\cE^a_\La] = 0.
\ee
Due to the identity $(\dl\circ\dl)(L)=0$,
$\eta(\dr_B\cE_A)^\La=(-1)^{[A][B]}\dr_A^\La\cE_B$, we obtain
\be
&& \op\sum_{0\leq|\La|}(-1)^{[a]}[(-1)^{[b]([a]+1)}\dr^{\La
a}\cE_b\cE_{\La
a} + (-1)^{[b][a]}\dr_a^\La\cE_b\cE^a_\La]=0, \\
&& \op\sum_{0\leq|\La|}(-1)^{[a]+1}[(-1)^{([b]+1)([a]+1)}\dr^{\La
a}\cE^b\cE_{\La a} + (-1)^{([b]+1)[a]}\dr_a^\La\cE^b\cE^a_\La]=0,
\ee
for all $\cE_b$ and $\cE^b$. This is exactly condition (iv).
\end{proof}

The equality (\ref{w44}) is called the master equation. For
instance, any variationally trivial Lagrangian satisfies the
master equation.

Being an element of the DGA $P^*_\infty\{N\}$ (\ref{w6}), an
original Lagrangian $L$ obeys the master equation (\ref{w44}) and
yields the graded derivations $\up_L=0$, $\ol\up_L=\ol\dl$
(\ref{w37}). The graded derivations (\ref{w37}) associated to the
Lagrangian $L_e$ (\ref{w8}) are extensions
\be
\up_e= u_e+ \frac{\op\dl^\lto \cL^*_1}{\dl \ol s_A}\frac{\dr}{\dr
s^A} + \op\sum_{0\leq k\leq N} \frac{\op\dl^\lto \cL^*_1}{\dl \ol
c_{r_k}}\frac{\dr}{\dr c^{r_k}}, \qquad
 \ol\up_e= \dl_{KT} +
\frac{\rdr }{\dr \ol s_A}\frac{\dl \cL_1}{\dl s^A}
\ee
of the gauge and KT operators, respectively. However, $L_e$ need
not satisfy the master equation. Therefore, let us consider its
extension
\mar{w61}\beq
L_E=L_e+L'=L+L_1+L_2+\cdots \label{w61}
\eeq
by means of even densities $L_i$, $i\geq 2$, of zero antifield
number and polynomial degree $i$ in ghosts. The corresponding
graded derivations (\ref{w37}) read
\mar{w102,3}\ben
&& \up_E= \up_e+ \frac{\op\dl^\lto \cL'}{\dl \ol
s_A}\frac{\dr}{\dr s^A} + \op\sum_{0\leq k\leq N}
\frac{\op\dl^\lto \cL'}{\dl \ol
c_{r_k}}\frac{\dr}{\dr c^{r_k}}, \label{w102} \\
&& \ol\up_E= \ol\up_e + \frac{\rdr }{\dr \ol
s_A}\frac{\dl\cL'}{\dl s^A} + \op\sum_{0\leq k\leq N} \frac{\rdr
}{\dr \ol c_{r_k}}\frac{\dl \cL'}{\dl c^{r_k}}. \label{w103}
\een
A Lagrangian $L_E$ (\ref{w61}) where $L+L_1=L_e$ is called a
proper extension of an original Lagrangian $L$. The following is a
corollary of Theorem \ref{w39}.

\begin{cor} \label{w120} \mar{w120} A Lagrangian $L$ is extended
to a proper  solution $L_E$ (\ref{w61}) of the master equation
only if the gauge operator $u_e$ (\ref{w108}) admits a nilpotent
extension.
\end{cor}

By virtue of condition (iv) of Theorem \ref{w39}, this nilpotent
extension is the derivation $\vt_E=\up_E +\ol\up^l_E$
(\ref{lmp1}), called the KT-BRST operator. With this operator, the
module of densities $P^{0,n}_\infty\{N\}$ is split into the
KT-BRST complex
\mar{lmp12}\beq
\cdots\to P^{0,n}_\infty\{N\}_2\to P^{0,n}_\infty\{N\}_1\to
P^{0,n}_\infty\{N\}_0\to P^{0,n}_\infty\{N\}^1\to
P^{0,n}_\infty\{N\}^2\to\cdots. \label{lmp12}
\eeq
Putting all ghosts zero, we obtain a cochain morphism of this
complex onto the KT complex, extended to
$\ol\cP^{0,n}_\infty\{N\}$ and reversed into the cochain one.
Letting all antifields zero, we come to a cochain morphism of the
KT-BRST complex (\ref{lmp12}) onto the cochain sequence
(\ref{w108}), where the gauge operator is extended to the
antifield-free part of the KT-BRST operator.

\begin{theo} \label{w130} \mar{w130} If the gauge
operator $u_e$ (\ref{w108}) can be extended to a nilpotent graded
derivation
\mar{w109}\beq
u_E=u_e+ \xi= u^A\dr_A + \op\sum_{1\leq k\leq N}(u^{r_{k-1}}
+\xi^{r_{k-1}})\dr_{r_{k-1}}+\xi^{r_N}\dr_{r_N} \label{w109}
\eeq
of ghost number 1 by means of antifield-free terms of higher
polynomial degree in ghosts, then the master equation has a proper
solution
\mar{w133}\beq
L_E=L_e + \op\sum_{1\leq k\leq N}\xi^{r_{k-1}}\ol c_{r_{k-1}}
d^nx= L+u_E( \op\sum_{0\leq k\leq N} c^{r_{k-1}}\ol c_{r_{k-1}})
d^nx +d_H\si. \label{w133}
\eeq
\end{theo}

\begin{proof} It is easily justified that, if the graded derivation
$u_E$ (\ref{w109}) is nilpotent, then  the right hand sides of the
equalities (\ref{w34}) equal zero. Using the relations
(\ref{qq1a}) -- (\ref{qq1c}), one can show that, in this case, the
right hand sides of the higher-stage NI (\ref{v93}) also equal
zero \cite{jmp05}. It follows that the summand $G_{r_k}$ of each
cocycle $\Delta_{r_k}$ (\ref{v92'}) is $\dl_{k-1}$-closed. Then
its summand $h_{r_k}$ is also $\dl_{k-1}$-closed and,
consequently, $\dl_{k-2}$-closed. Hence it is $\dl_{k-1}$-exact by
virtue of the homology regularity condition. Therefore,
$\Delta_{r_k}$ contains only the term $G_{r_k}$ linear in
antifields. It follows that the Lagrangian $L_e$ (\ref{w8}) and,
consequently, the Lagrangian $L_E$ (\ref{w133}) are affine in
antifields. In this case, we have $u^A=\op\dl^\lto{}^A(\cL_e)$,
$u^{r_k}=\op\dl^\lto{}^{r_k}(\cL_e)$ for all indices $A$ and $r_k$
and, consequently, $u^A_E=\op\dl^\lto{}^A(\cL_E)$,
$u^{r_k}_E=\op\dl^\lto{}^{r_k}(\cL_E)$, i.e., $u_E=\up_E$ is the
graded derivation (\ref{w102}) defined by the Lagrangian $L_E$.
Its nilpotency condition takes the form
$u_E(\op\dl^\lto{}^A(\cL_E))=0$,
$u_E(\op\dl^\lto{}^{r_k}(\cL_E))=0$. Hence, we obtain
\be
u_E(\cL_E)=u_E(\op\dl^\lto{}^A(\cL_E)\ol s_A +
\op\dl^\lto{}^{r_k}(\cL_E) \ol c_{r_k})=0,
\ee
i.e., $u_E$ is a variational symmetry of $L_E$. Consequently,
$L_E$ obeys the master equation.
\end{proof}

\begin{cor} \label{lmp6} \mar{lmp6} The gauge operator
(\ref{w108}) admits the nilpotent extension (\ref{w109}) only if
gauge symmetry conditions (\ref{w34}) and the higher-stage NI
(\ref{v93}) are satisfied off-shell. In this case, the Lagrangian
$L_e$ (\ref{w8}) is affine in antifields.
\end{cor}

The nilpotent extension (\ref{w109}) of the gauge operator is
called the BRST operator. It brings the cochain sequence
(\ref{w108}) into the BRST complex.

By virtue of Corollary \ref{lmp6}, the gauge operator $u_e$ is
extended to the BRST operator only if gauge symmetry conditions
hold off-shell, i.e., $u_e(u_e)=u(u_e)$. Let us consider a
particular case when higher-stage gauge symmetries $u_{(k)}$
(\ref{w38}) are independent of original fields $s^A$, i.e.,
$u_e(u_e)=u(u)$. In this case, the BRST operator $u_E$
(\ref{w109}) reads
\mar{lmp15}\beq
u_E=u_e+\xi=u_e +\xi^r\dr_r, \quad (u+\xi)(u^A)=0, \quad (u +
\xi)(\xi^r)=0, \quad u_{(1)}(\xi^r)=0. \label{lmp15}
\eeq
For instance, irreducible theories and Abelian reducible theories,
where $u(u)=0$, are of this type. One can think of the first and
second conditions (\ref{lmp15}) as being the generalized
commutation relations and the Jacobi identity, respectively
\cite{jmp05}. It follows from the second one that $\xi$ is
quadratic in ghosts. Moreover, if a gauge symmetry $u$ is
polynomial in fields, then it is necessarily affine, and $\xi$ is
independent of original fields. In Abelian reducible theories, the
gauge operator $u_e$ itself is nilpotent. The corresponding proper
solution (\ref{w133}) of the master equation  reads
\mar{lmp30}\beq
L_E=L +\dl_{KT}( \op\sum_{0\leq k\leq N} c^{r_k}\ol c_{r_k}d^nx) =
L+u_e( \op\sum_{0\leq k\leq N} c^{r_{k-1}}\ol c_{r_{k-1}})d^nx +
d_H\si. \label{lmp30}
\eeq

\section{Examples}

1) Yang--Mills gauge theory exemplifies an irreducible degenerate
Lagrangian theory where $u_E$ (\ref{w109}) is the familiar BRST
operator \cite{gom}. Let us consider Yang--Mills supergauge theory
where gauge symmetries form a finite-dimensional Lie superalgebra
over $C^\infty(X)$. Let $\cG=\cG^0\oplus \cG^1$ be a real Lie
superalgebra with a basis $\{e_r\}$, $r=1,\ldots,m,$ and real
structure constants $c^r_{ij}$. We denote $[e_r]=[r]$. Given the
enveloping algebra $\ol \cG$ of $\cG$, let us assume that there is
an invariant even element $h^{ij}e_ie_j$ of $\ol\cG$ such that the
matrix $h^{ij}$ is non-degenerate. The gauge theory of $\cG$ on
$X=\Bbb R^n$ is described by the DGA $\cS^*[F;Y]$ where
\be
F=(X\times \cG)\ot T^*X\to (X\times \cG^0)\ot T^*X\to X.
\ee
Its basis is $(a^r_\la)$, $[a^r_\la]=[r]$. The Yang--Mills
Lagrangian reads
\be
L_{YM}=\frac14
h_{ij}\eta^{\la\m}\eta^{\bt\nu}\cF^i_{\la\bt}\cF^j_{\m\nu}d^nx,
\qquad \cF^r_{\la\m} =a^r_{\la\m}-a^r_{\m\la} +c^r_{ij}a^i_\la
a^j_\m.
\ee
Its Euler--Lagrange operator obeys the irreducible NI
\be
c^r_{ji}a^i_\la\cE_r^\la + d_\la\cE_j^\la=0.
\ee
Therefore, we enlarge $\cS^*[F;Y]$ to the DGA
\be
P^*\{0\}=\cP^*[\ol{VF}\op\oplus_Y E_0\op\oplus_Y \ol E_0;Y],
\qquad E_0=X\times \cG_0,
\ee
whose basis $(a^r_\la, c^r, \ol a^\la_r, \ol c_r)$, $
[c^r]=([r]+1){\rm mod}\,2$, $[\ol a^\la_r]=[\ol c_r]=[r]$,
contains ghosts $c^r$ of ghost number 1 and antifields $\ol
a^\la_r$, $\ol c_r$ of antifield numbers 1 and 2, respectively.
Then the gauge operator $u_e$ (\ref{w108}) and the Lagrangian
$L_e$ (\ref{w8}) read
\be
u_e= (-c^r_{ji}c^ja^i_\la + c^r_\la)\frac{\dr}{\dr a_\la^r},\qquad
 L_e=  L_{YM} + (-c^r_{ji}c^ja^i_\la + c^r_\la)\ol
a^\la_r d^nx.
\ee
The gauge operator $u_e$ admits the BRST extension
\be
u_E=u_e +\xi= (-c^r_{ji}c^ja^i_\la + c^r_\la)\frac{\dr}{\dr
a_\la^r} -\frac12 (-1)^{[i]} c^r_{ij}c^ic^j\frac{\dr}{\dr c^r}.
\ee
Then the proper solution (\ref{w133}) of the master equation takes
the form
\be
L_E=L_{YM}+ (-c^r_{ij}c^ja^i_\la + c^r_\la)\ol a^\la_rd^nx
-\frac12 (-1)^{[i]} c^r_{ij}c^ic^j\ol c_r d^nx.
\ee

2) In contrast with Yang--Mills gauge theory, gauge symmetries of
gravitation theory fail to form a finite-dimensional Lie algebra.
Gravitation theory can be formulated as gauge theory on natural
bundles over a four-dimensional manifold $X$. These bundles admit
the functorial lift $\wt\tau$ of any vector field $\tau$ on $X$
such that $\tau\mapsto\ol\tau$ is a Lie algebra monomorphism. This
lift is an infinitesimal generator of a local one-parameter group
of general covariant transformations. Dynamic variables of gauge
gravitation theory are linear connections and pseudo-Riemannian
metrics on $X$. The first ones are principal connections on the
linear frame bundle $LX$ of $X$. They are represented by sections
of the bundle $C_K=J^1LX/GL(4,\Bbb R)$. Pseudo-Riemannian metrics
on $X$ are sections of the quotient bundle $\Si=LX/O(1,3)$. The
total configuration space of gravitation theory $\Si\times C_K$,
coordinated by $(x^\la,\si^{\al\bt}, k_\mu{}^\al{}_\bt)$, admits
the functorial lift
\be
&&\tau +(\si^{\nu\bt}\dr_\nu \tau^\al
+\si^{\al\nu}\dr_\nu \tau^\bt)\frac{\dr}{\dr \si^{\al\bt}}\\
&&\qquad +
(\dr_\nu \tau^\al k_\m{}^\nu{}_\bt -\dr_\bt \tau^\nu
k_\m{}^\al{}_\nu -\dr_\mu \tau^\nu k_\nu{}^\al{}_\bt
+\dr_{\m\bt}\tau^\al)\frac{\dr}{\dr k_\mu{}^\al{}_\bt}
\ee
of vector fields $\tau$ on $X$ \cite{book00}. Let a gravitation
Lagrangian $L_{MA}$ on the jet manifold $J^1(\Si\times C_K)$ be
invariant under general covariant transformations. Then the
variational derivatives $\cE_{\al\bt}$, $\cE^\m{}_\al{}^\bt$ of
this Lagrangian obey the irreducible NI
\be
&&-(\si^{\al\bt}_\la +2\si^{\nu\bt}_\nu\dl^\al_\la)\cE_{\al\bt}
-2\si^{\nu\bt}d_\nu\cE_{\la\bt} +(-k_{\la\m}{}^\al{}_\bt
-k_{\nu\m}{}^\nu{}_\bt\dl^\al_\la + k_{\bt\m}{}^\al{}_\la \\
&& \qquad +
k_{\m\la}{}^\al{}_\bt)\cE^\m{}_\al{}^\bt
+(-k_\m{}^\nu{}_\bt\dl^\al_\la +k_\m{}^\al{}_\la\dl^\nu_\bt
+k_\la{}^\al{}_\bt\dl^\nu_\m)d_\nu\cE^\m{}_\al{}^\bt + d_{\m\bt}
\cE^\m{}_\la{}^\bt=0.
\ee
Taking the vertical part of vector fields $\wt\tau$ and replacing
gauge parameters $\tau^\la$ with ghosts $c^\la$, we obtain the
gauge operator $u_e$ and its nilpotent BRST prolongation
\be
&&u_E=(\si^{\nu\bt} c_\nu^\al +\si^{\al\nu}
c_\nu^\bt-c^\la\si_\la^{\al\bt})\frac{\dr}{\dr \si^{\al\bt}}+
\\
&& \qquad (c_\nu^\al k_\m{}^\nu{}_\bt -c_\bt^\nu k_\m{}^\al{}_\nu
-c_\mu^\nu k_\nu{}^\al{}_\bt +c_{\m\bt}^\al-c^\la
k_{\la\mu}{}^\al{}_\bt)\frac{\dr}{\dr k_\mu{}^\al{}_\bt} +
c^\la_\m c^\m\frac{\dr}{\dr c^\la}.
\ee
Accordingly, an original gravitation Lagrangian $L_{MA}$ is
extended to a proper solution
\be
&& L_E= L_{MA} + (\si^{\nu\bt} c_\nu^\al +\si^{\al\nu}
c_\nu^\bt-c^\la\si_\la^{\al\bt}) \ol\si_{\al\bt} d^nx + \\
&& \qquad (c_\nu^\al k_\m{}^\nu{}_\bt -c_\bt^\nu k_\m{}^\al{}_\nu
-c_\mu^\nu k_\nu{}^\al{}_\bt +c_{\m\bt}^\al-c^\la
k_{\la\mu}{}^\al{}_\bt) \ol k^\m{}_\al{}^\bt d^nx + c^\la_\m c^\m
\ol c_\la d^nx
\ee
of the master equation where $\ol\si_{\al\bt}$, $\ol
k^\m{}_\al{}^\bt$ and $\ol c_\la$ are the corresponding
antifields.

3) Topological BF theory exemplifies an Abelian reducible
degenerate Lagrangian theory which satisfies the homology
regularity condition \cite{jpa05,jmp05a}. Its dynamic variables
are exterior forms $A$ and $B$ of form degree $|A| +|B|=n-1$ on a
manifold $X$. They are sections of the bundle $Y=\op\w^pT^*X\oplus
\op\w^qT^*X$, $p+q=n-1$, coordinated by $(x^\la,
A_{\m_1\ldots\m_p},B_{\nu_1\ldots\nu_q})$. Without a loss of
generality, let $q$ be even and $q\geq p$. The corresponding DGA
is $\cO^*_\infty Y$. There are the canonical $p$- and $q$-forms
\be
A=\frac{1}{p!}A_{\m_1\ldots\m_p}dx^{\m_1}\w\cdots\w dx^{\m_p},
\qquad
B=\frac{1}{q!}B_{\nu_{p+1}\ldots\nu_q}dx^{\nu_{p+1}}\w\cdots\w
dx^{\nu_p}
\ee
on $Y$. A Lagrangian  $L_{\rm BF}=A\w d_HB$ leads to
Euler--Lagrange equations $d_HA=0$, $d_HB=0$ obeying NI $d_Hd_HA=
0$, $d_Hd_H B= 0$. Given the vector bundles
\be
&& E_k=\op\w^{p-k-1}T^*X\op\times_X \op\w^{q-k-1}T^*X, \quad 0\leq
k< p-1, \qquad  E_{p-1}=\Bbb R \op\times_X
\op\w^{q-p}T^*X,  \\
&& E_k=\op\w^{q-k-1}T^*X, \quad p-1<k<q-1, \qquad E_{q-1}=X\times
\Bbb R,
\ee
let us consider the DGA $P_\infty^*\{q-1\}$ with a local basis
\be
&& \{A_{\m_1\ldots\m_p}, B_{\nu_{p+1}\ldots\nu_q},
\ve_{\m_2\ldots\m_p},\ldots,\ve_{\m_p},\ve,\xi_{\nu_{p+2}\ldots\nu_q},
\ldots, \xi_{\nu_q},\xi,\\
&&\qquad \ol A^{\m_1\ldots\m_p}, \ol B^{\nu_{p+1}\ldots\nu_q},
\ol\ve^{\m_2\ldots\m_p}, \ldots,\ol\ve^{\m_p}, \ol \ve, \ol
\xi^{\nu_{p+2}\ldots\nu_q}, \ldots, \ol \xi^{\nu_q},\ol \xi\}.
\ee
Then the gauge operator $u_e$ (\ref{w108}) reads
\be
&& u_e= d_{\m_1}\ve_{\m_2\ldots\m_p}\frac{\dr}{\dr
A_{\m_1\m_2\ldots\m_p}} +
d_{\nu_{p+1}}\xi_{\nu_{p+2}\ldots\nu_q}\frac{\dr}{\dr
B_{\nu_{p+1}\nu_{p+2}\ldots\nu_q}}\\
&& \qquad + [d_{\m_2}\ve_{\m_3\ldots\m_p}\frac{\dr}{\dr
\ve_{\m_2\m_3\ldots\m_p}}+\cdots  +d_{\m_p}\ve\frac{\dr}{\dr
\ve^{\m_p}}]\\
&&\qquad + [d_{\nu_{p+2}}\xi_{\nu_{p+3}\ldots\nu_q}
\frac{\dr}{\dr \xi_{\nu_{p+2}\nu_{p+3}\ldots\nu_q}}+\cdots +
d_{\nu_q}\xi\frac{\dr}{\dr \xi^{\nu_q}}].
\ee
This operator is obviously nilpotent and, thus, is the BRST
operator. Consequently, the proper extension of the Lagrangian
$L_{\rm BF}$ takes the form (\ref{lmp30}):
\be
&& L_E=L_{\rm BF} + [\ve_{\m_2\ldots \m_p} d_{\m_1} \ol
A^{\m_1\m_2\ldots \m_p} + \ve_{\m_3\ldots \m_p} d_{\m_2} \ol
\ve^{\m_2\m_3\ldots \m_p} +\cdots +\ve d_{\m_p}\ol \ve^{\m_p}] d^nx+\\
&& \qquad [\xi_{\nu_{p+2}\ldots \nu_q} d_{\nu_{p+1}} \ol
A^{\nu_{p+1}\nu_{p+2}\ldots \nu_q} + \xi_{\nu_{p+3}\ldots \nu_q}
d_{\nu_{p+2}} \ol \ve^{\nu_{p+2}\nu_{p+3}\ldots \nu_q} +\cdots
+\xi d_{\nu_q}\ol \ve^{\nu_q}] d^nx.
\ee

\end{document}